\documentclass[journal,twoside,web,dvipsnames]{ieeecolor}          %
\usepackage{etoolbox}
\makeatletter
\@ifundefined{color@begingroup}%
  {\let\color@begingroup\relax
   \let\color@endgroup\relax}{}%
\def\fix@ieeecolor@hbox#1{%
  \hbox{\color@begingroup#1\color@endgroup}}
\patchcmd\@makecaption{\hbox}{\fix@ieeecolor@hbox}{}{\FAILED}
\patchcmd\@makecaption{\hbox}{\fix@ieeecolor@hbox}{}{\FAILED}

\newcommand{\usetmi}{}

\usepackage{tmi}
\usepackage{cite}
\usepackage{amsmath,amssymb,amsfonts}
\usepackage{graphicx}
\usepackage{textcomp}
\usepackage[most]{tcolorbox}
\usepackage{tikz}
\usepackage[customcolors]{hf-tikz}
\usepackage{algorithm}
\usepackage[noend]{algpseudocode}
\usepackage{siunitx}
\usepackage{subfig}
\usepackage{url}
\usepackage{xcolor}  %
\usepackage{mathtools}
\usepackage{flushend}
\usepackage[hidelinks]{hyperref}

\colorlet{Lavender}{Lavender!80!Black}
\definecolor{TMIBlue}{RGB}{0, 138, 218}
\definecolor{IEEEBlue}{RGB}{0, 67, 147}
\definecolor{ReviewAmber}{RGB}{204, 153, 0}

\definecolor{IEEEBlueFill}{RGB}{0,112,192}     %
\definecolor{IEEEBlueBorder}{RGB}{0,64,128}    %
\definecolor{GrayFill}{gray}{0.7}
\definecolor{GrayBorder}{gray}{0.3}

\DeclareSIUnit{\nothing}{\relax}

\tikzset{
  style green/.style={
    set fill color=ForestGreen,
    fill opacity=0.4,
    draw=none,
  },
  style lavender/.style={
    set fill color=Lavender,
    fill opacity=0.4,
    draw=none,
  },
  style blue/.style={
    set fill color=TMIBlue,
    fill opacity=0.4,
    draw=none,
  },
  ver/.style={
    above left offset={-0.2,0.3},
    below right offset={0.1,-0.2},
    #1
  },
  hor/.style={
    above left offset={-0.3,0.35},
    below right offset={0.3,-0.3},
    #1
  }
}

\newcommand{\tsw}[1]{\textcolor{black}{#1}}

\newcommand{\redcircle}{\tikz[baseline=-0.6ex] \draw[red, fill=red, fill opacity=0.3] (0,0) circle (3pt);}

\newcommand{\bluecircle}{\tikz[baseline=-0.6ex]\draw[fill=IEEEBlueFill, draw=IEEEBlueBorder] (0,0) circle (3pt);}
\newcommand{\graycircle}{\tikz[baseline=-0.6ex]\draw[fill=GrayFill, draw=GrayBorder] (0,0) circle (3pt);}

\newcommand{\bluecross}{\tikz[baseline=-0.6ex]\node[inner sep=0pt, outer sep=0pt, text=TMIBlue] at (0,0) {\small$\times$};}
\newcommand{\graycross}{\tikz[baseline=-0.6ex]\node[inner sep=0pt, outer sep=0pt, text=Gray] at (0,0) {\small$\times$};}

\usepackage[acronym]{glossaries}
\newacronym{awgn}{AWGN}{additive white Gaussian noise}
\newacronym{dgm}{DGM}{deep generative model}
\newacronym{dps}{DPS}{diffusion posterior sampling}
\newacronym{dm}{DM}{diffusion model}
\newacronym{lpips}{LPIPS}{learned perceptual image patch similarity}
\newacronym{psnr}{PSNR}{peak signal-to-noise ratio}
\newacronym{ef}{EF}{ejection fraction}
\newacronym{snr}{SNR}{signal-to-noise ratio}
\newacronym{mi}{MI}{mechanical index}
\newacronym{gan}{GAN}{generative adversarial network}
\newacronym{vae}{VAE}{variational autoencoder}
\newacronym{mmse}{MMSE}{minimum mean square error estimator}
\newacronym{ema}{EMA}{exponential moving average}
\newacronym{seqdiff}{SeqDiff}{sequential diffusion}
\newacronym{cs}{CS}{compressed sensing}
\newacronym{rf}{RF}{radio-frequency}
\newacronym{ood}{OoD}{out-of-distribution}
\newacronym{cdm}{CDM}{conditional diffusion model}
\newacronym{inr}{INR}{implicit neural representation}
\newacronym{ssim}{SSIM}{structural similarity index measure}

\usepackage{amsmath,amsfonts,bm}

\def\Secref#1{Section~\ref{#1}}

\def\1{\bm{1}}

\def\rvepsilon{{\mathbf{\epsilon}}}

\def\rvn{{\mathbf{n}}}

\def\rvx{{\mathbf{x}}}
\def\rvy{{\mathbf{y}}}

\def\rmI{{\mathbf{I}}}

\def\mA{{\bm{A}}}

\def\mI{{\bm{I}}}

\def\mM{{\bm{M}}}

\DeclareMathAlphabet{\mathsfit}{\encodingdefault}{\sfdefault}{m}{sl}
\SetMathAlphabet{\mathsfit}{bold}{\encodingdefault}{\sfdefault}{bx}{n}
\newcommand{\tens}[1]{\bm{\mathsfit{#1}}}

\def\tE{{\tens{E}}}

\def\tI{{\tens{I}}}

\def\tX{{\tens{X}}}
\def\tY{{\tens{Y}}}

\def\sR{{\mathbb{R}}}

\newcommand{\E}{\mathbb{E}}

\newcommand{\R}{\mathbb{R}}

\newcommand{\norm}[1]{\left \lVert #1 \right \rVert}

\newcommand{\defeq}{\vcentcolon=}

\def\BibTeX{{\rm B\kern-.05em{\sc i\kern-.025em b}\kern-.08em
    T\kern-.1667em\lower.7ex\hbox{E}\kern-.125emX}}

\begin{document}
\title{High Volume Rate 3D Ultrasound Reconstruction with Diffusion Models}
\author{Tristan S.W. Stevens, \IEEEmembership{Graduate Student Member, IEEE}, Oisín Nolan, \IEEEmembership{Graduate Student Member, IEEE}, \\ Oudom Somphone, Jean-Luc Robert and Ruud J.G. van Sloun, \IEEEmembership{Member, IEEE} \vspace{-1cm}
\thanks{
Manuscript received 7 July 2025; revised 6 November 2025; accepted 13 December 2025. This work was performed within the IMPULSE framework of the Eindhoven MedTech Innovation Center (e/MTIC, incorporating Eindhoven University of Technology and Philips Research), including a PPS supplement from the Dutch Ministry of Economic Affairs and Climate Policy.\\
\indent Tristan S.W. Stevens (e-mail: t.s.w.stevens@tue.nl), Oisín Nolan and Ruud J.G. van Sloun are with the Electrical Engineering Department,
Eindhoven University of Technology, the Netherlands. Oudom Somphone and Jean-Luc Robert are with Philips Research, Paris, France and Cambridge MA, USA, respectively.}}

\bibliographystyle{IEEEtran}

\maketitle
\begin{abstract}
Three-dimensional ultrasound enables real-time volumetric visualization of anatomical structures. Unlike traditional 2D ultrasound, 3D imaging reduces reliance on precise probe orientation, potentially making ultrasound more accessible to clinicians with varying levels of experience and improving automated measurements and post-exam analysis. However, achieving both high volume rates and high image quality remains a significant challenge. While 3D diverging waves can provide high volume rates, they suffer from limited tissue harmonic generation and increased multipath effects, which degrade image quality. One compromise is to retain focus in elevation while leveraging unfocused diverging waves in the lateral direction to reduce the number of transmissions per elevation plane. Reaching the volume rates achieved by full 3D diverging waves, however, requires dramatically undersampling the number of elevation planes. Subsequently, to render the full volume, simple interpolation techniques are applied. This paper introduces a novel approach to 3D ultrasound reconstruction from a reduced set of elevation planes by employing diffusion models (DMs) to achieve increased spatial and temporal resolution. We compare both traditional and supervised deep learning-based interpolation methods on a 3D cardiac ultrasound dataset. Our results show that DM-based reconstruction consistently outperforms the baselines in image quality and downstream task performance. Additionally, we accelerate inference by leveraging the temporal consistency inherent to ultrasound sequences. Finally, we explore the robustness of the proposed method by exploiting the probabilistic nature of diffusion posterior sampling to quantify reconstruction uncertainty and demonstrate improved recall on out-of-distribution data with synthetic anomalies under strong subsampling. Code is available at \href{http://3d-ultrasound-diffusion.github.io}{3d-ultrasound-diffusion.github.io}
\end{abstract}

\begin{IEEEkeywords}
3D ultrasound, diffusion models, generative modeling, cardiac ultrasound
\end{IEEEkeywords}

\section{Introduction}
\label{sec:introduction}
\IEEEPARstart{T}{hree}-dimensional (3D) ultrasound imaging, also known as volumetric ultrasound imaging, is achieved through the use of a 2D matrix probe which enables focusing of ultrasound beams in both azimuth and elevation dimensions through electronic steering \cite{stetten1998real}. It overcomes several challenges associated with traditional 2D ultrasound imaging, such as the inability to capture out-of-plane motion and the lack of spatial orientation. These limitations make 2D imaging highly operator-dependent, requiring multiple scans to mentally reconstruct anatomy, which is time-consuming, variable across users, and difficult to reproduce.  As a result, diagnostic accuracy and interventional guidance can suffer from inconsistency~\cite{nelson1998three, fenster2011three}.

The transition to 3D ultrasound imaging addresses these limitations by offering volumetric data, enabling applications such as 3D echocardiography~\cite{pedrosa2016cardiac, wu2017three, papademetris19993d} and 3D breast imaging~\cite{catalano2023recent}.
It also provides improved guidance during image-guided therapy and biopsies~\cite{fenster20023d} and supports greater automation of critical diagnostic measurements, including ejection fraction~\cite{spitzer2017role} and strain assessment~\cite{de20133d, de20253}.

However, 3D imaging introduces new challenges, such as increased data acquisition times and higher computational demands \cite{huang2017review}. These factors constrain the design of probes and transmit sequences, limiting the accessibility of high-quality 3D imaging in real-time clinical settings. Additionally, the reduced image quality compared to 2D imaging often discourages clinicians from utilizing 3D imaging, as the degradation in resolution and clarity frequently outweighs its potential benefits in practical scenarios.

In addressing the trade-offs inherent to 3D ultrasound, it is beneficial to differentiate between two key phases of the imaging process: \emph{acquisition} and \emph{reconstruction}~\cite{van2024active}. Acquisition refers to the process of transmitting and receiving ultrasound waves, which can be optimized to minimize data acquisition times while maintaining sufficient information for effective image reconstruction. The increased field of view in 3D imaging, while enabling more comprehensive anatomical coverage, leads to acquisition times that are prohibitively long under standard transmit schemes. Combined with the large number of elements in 2D matrix arrays, which significantly increases data rates, this has necessitated the development of bandwidth-efficient solutions~\cite{huang2017review}.

Microbeamforming reduces data transmission by performing partial beamforming within the probe itself, grouping elements into sub-apertures, at the cost of reduced lateral resolution~\cite{castrignano2025impact}. Slow-time multiplexing limits the number of active elements per transmission, reducing bandwidth but lowering frame rates \cite{rezvanitabar2022integrated}. Other techniques like \acrfull{cs} aim to reduce the number of samples needed by exploiting the inherent sparsity of ultrasound signals. In 2D ultrasound, Wagner~\emph{et~al.} \cite{wagner2012compressed} applied \acrshort{cs} to \acrfull{rf} signals using Xampling, simplifying acquisition through structured signal representations. This was extended to 3D ultrasound by Burshtein et al. \cite{burshtein2016sub}, who demonstrated the approach for volumetric imaging. Learned approaches for designing sparse sensing matrices in ultrasound imaging have shown promising results. Lorintiu~\emph{et~al.} \cite{lorintiu2015compressed} employed learned overcomplete dictionaries for scanline selection in volumetric ultrasound, while Huijben~\emph{et~al.} \cite{huijben2020learning} proposed a deep learning-based probabilistic subsampling scheme that generates context-specific sensing matrices for 2D ultrasound.

Reconstruction, on the other hand, involves the process of generating a 3D image or volume from the acquired data. Traditional reconstruction techniques, such as pixel or voxel nearest-neighbor (PNN or VNN) interpolation, simply fill in missing information by relying on nearby sampled points~\cite{ji2011real, huang2017review}. Function-based interpolation methods are an improved, albeit usually more computationally demanding, alternative, where a particular function, for example, a polynomial, is fitted through the acquired voxels.

Recently, deep learning has emerged as a powerful tool for ultrasound image reconstruction \cite{van2019deep}. Specifically, \acrfullpl{dgm} have proven to be the missing link in learning expressive image priors \cite{van2024active, stevensDeepGenerative2025}.
Unlike conventional supervised approaches that learn a direct mapping from measurements to images, generative models learn the underlying data distribution in a self-supervised manner. This enables reconstruction to be framed as \emph{Bayesian} inference, where the \emph{likelihood} captures the forward measurement process and the generative model acts as a learned \emph{prior}. \emph{Posterior} sampling then allows for the recovery of plausible and diverse image reconstructions from sparse or noisy inputs. Consequently, \acrshortpl{dgm} are task-agnostic and naturally support uncertainty estimation within a probabilistic framework, improving robustness to \acrfull{ood} data. A specific subset of \acrshortpl{dgm}, \acrfullpl{gan}, has been successfully applied in 3D ultrasound imaging to upscale sparsely acquired 2D images to reconstruct a full volume \cite{he2020deep, dai2021self}. Also, \acrfullpl{vae} have been used to synthesize 3D ultrasound data with the purpose of augmenting existing datasets \cite{wulff2023towards}. Beyond ultrasound, \acrshortpl{dgm} for volumetric imaging have also been studied in other modalities, including the synthesis of high-fidelity 3D CT and MRI volumes~\cite{Guo2025-sv,sun2022hierarchical,chen2024towards}.

More recently, diffusion models have seen a surge of interest due to their powerful generative capabilities and practical training objective. In the field of ultrasound, diffusion models are used for image generation \cite{stojanovski2023echo, dominguez2024diffusion, freiche2025ultrasound}, denoising and dehazing \cite{stevens_removing_2023, stevens2024dehazing, asgariandehkordi2024denoising}, and image reconstruction \cite{zhang2023ultrasound, durrer2024denoising, penninga2025deep}. While diffusion models generally converge more reliably during training compared to \acrshortpl{gan} and produce higher quality samples than \acrshortpl{vae}, they tend to be more computationally demanding during inference due to their iterative nature~\cite{cao2024survey}. Many works have shown, however, that it is feasible to accelerate diffusion models using various techniques, such as leveraging structure in sequential data \cite{stevens2024sequential} or model distillation \cite{meng2023distillation}.

Another consideration is the use of \acrfullpl{cdm}. While \acrshortpl{cdm} are popular for synthesizing medical images based on class labels~\cite{mishra2023dual}, their application to image reconstruction is uncommon, as it sacrifices several benefits of unconditional \acrshortpl{dm}, including being task agnostic and the ability to leverage general foundational models~\cite{jiao2024usfm}.

In this work, we propose a flexible interpolation framework based on diffusion models for 3D ultrasound reconstruction, enabling high-volume rates without compromising image quality. Specifically, we demonstrate our approach on an \emph{in-vivo} cardiac 3D ultrasound dataset, showing that reliable reconstructions are achievable at $\mathbf{3\times}$ increased volume rate while maintaining image quality and downstream task performance.

Our main contributions can be summarized as follows:
\begin{itemize}
    \item Development of a deep generative prior for 3D cardiac ultrasound in the form of a diffusion model.
    \item A practical posterior sampling framework that can interpolate sparsely sampled ultrasound volumes.
    \item Techniques to mitigate and visualize uncertainty of the generative model by drawing multiple posterior samples.
    \item An extensive comparison with existing interpolation techniques, showcasing the advantage of using deep generative priors, with evaluation based on both image quality metrics and a clinically relevant downstream task.
\end{itemize}

\begin{figure}
    \centering
    \includegraphics[width=0.8\linewidth]{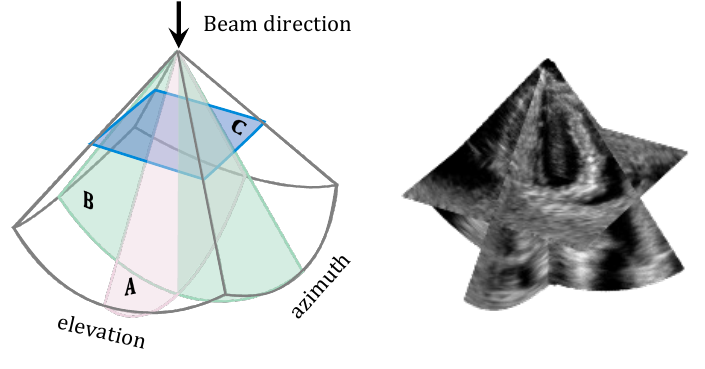}
    \vspace{-0.2cm}
    \caption{Geometric visualization of the three common cross-sections in volumetric ultrasound imaging (left) with a B-mode example (right).}
    \label{fig:planes}
\end{figure}

\section{Background}
In this section, we will briefly discuss basic concepts in 3D ultrasound imaging and diffusion models and their use in learning image priors, which form the basis of the proposed methodology.

\subsection{3D Ultrasound acquisition}\label{sec:3d-ultrasound}
Modern ultrasound probes that enable 3D imaging form a matrix of transducer elements, that allows for dynamic beam steering along both azimuth (lateral) and elevation axes. The fully acquired 3D ultrasound volume $\tX\in\sR^{N_\text{el}, N_\text{az}, N_\text{ax}}$, after beamforming, envelope-detection and log-compression, can be written as a tensor:
\begin{equation}
    \tX =
    \left[\begin{array}{@{\hspace{0.8em}}c@{\hspace{0.5em}}c@{\hspace{0.5em}}c@{\hspace{1em}}c@{\hspace{0.8em}}}
        \tikzmarkin[ver=style green]{firstcol}\tikzmarkin[hor=style lavender]{firstrow}\rvx_{1,1} & \rvx_{1,2} & \cdots & \rvx_{1,N_\text{az}}\tikzmarkend{firstrow} \\[0.4em]
        \rvx_{2,1} & \rvx_{2,2} & \cdots & \rvx_{2,N_\text{az}} \\
        \vdots & \vdots & \ddots & \vdots \\[0.4em]
        \rvx_{N_\text{el},1}\tikzmarkend{firstcol} & \rvx_{N_\text{el},2} & \cdots & \rvx_{N_\text{el},N_\text{az}}
        \end{array}\right],
\label{eq:ultrasound-volume}
\end{equation}
with elements $\rvx\in\sR^{N_\text{ax}}$. The dimensions $N_\text{el}$, $N_\text{az}$, and $N_\text{ax}$ represent the number of samples acquired along the elevation, azimuth, and axial directions, respectively. From $\tX$, three orthogonal planes can be identified: the \textcolor{Lavender}{A}, \textcolor{ForestGreen}{B}, and \textcolor{TMIBlue}{C} planes. Rows in $\tX$ can be considered lateral 2D images at a fixed elevation plane, akin to traditional 2D B-mode imaging (\textcolor{Lavender}{A plane}). The columns represent elevation images (\textcolor{ForestGreen}{B plane}) which are perpendicular to the A plane. If A plane images provide a long-axis view, the B plane offers the corresponding short-axis view. \textcolor{TMIBlue}{The C plane}, represents slices parallel to the probe surface, acquired at a fixed axial depth. See Fig.~\ref{fig:planes} for a visualization of all three planes in relation to each other.

\begin{figure}
    \centering
    \includegraphics[width=0.95\linewidth]{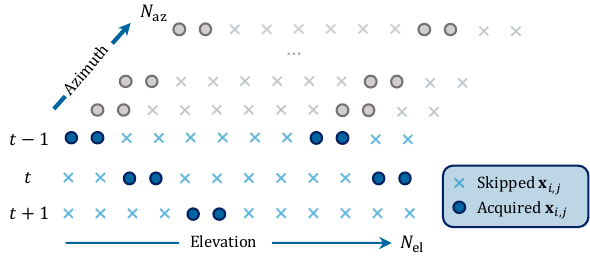}
    \caption{Schematic representation of the acquisition scheme with sparse interlocking patterns over time. The acquired \protect\bluecircle{} and skipped \protect\bluecross{} elevation planes at times $t{-}1$, $t$, and $t{+}1$ are marked in blue for the first azimuth line. This interleaved scheme ensures complementary elevation planes are captured across frames, supporting temporally consistent volumetric reconstruction. In gray (\protect\graycircle{}, \protect\graycross{}), the same acquisition pattern is repeated across other azimuth indices for context.}
    \label{fig:acquisition-scheme}
\end{figure}

In practice, probe design influences both image quality and acquisition strategies. For transthoracic transducers, the design is often asymmetric, featuring a smaller aperture along the elevation axis to facilitate access through narrow imaging windows between the ribs. The smaller aperture broadens the beam and reduces elevational resolution, while a lower element count with wider spacing increases grating lobe artifacts, degrading image quality. To address these limitations while maintaining sufficient temporal resolution, a faster acquisition scheme such as diverging waves is employed along the azimuth dimension, while focused transmits are applied along elevation at a coarse interval.
Employing only diverging waves in both dimensions significantly degrades image quality because the unfocused beams produce lower acoustic pressure, resulting in reduced harmonic generation and increased multipath scattering. To compensate for these effects, higher transmit power is required, which raises the \acrfull{mi} and poses challenges for transducer design and patient safety. In contrast, fully focusing across both dimensions improves image quality and resolution, but at the cost of frame rate, since more transmits are needed to cover the entire field of view. As a compromise, combining diverging waves in azimuth with focused transmits along elevation balances image quality and acquisition speed. Limiting the number of focused elevation angles directly impacts the maximum achievable volume rate, but introduces gaps in the volumetric data. To mitigate this, we adopt a sparse interlocking acquisition scheme, as illustrated in Fig.~\ref{fig:acquisition-scheme}, where different elevation planes are acquired at each time step in a staggered pattern. This ensures that across consecutive frames, complementary slices are captured, allowing temporal redundancy to aid in volumetric reconstruction. Still, at lower acquisition rates, high-quality interpolation remains a challenge.

\subsection{Diffusion models}\label{sec:diffusion}
Generative models seek to learn the data distribution $p(\rvx)$ of some random variable $\rvx$. Instead of explicitly modeling the distribution, which is often intractable given high-dimensional data, these generative models implicitly model their distribution using a generative process, i.e. producing samples from the underlying distribution $\rvx \sim p(\rvx)$. \Acrfullpl{dm} \cite{ho2020denoising, song2020score} are a class of generative models that define the generative process as the reversal of a corruption process, which transforms ${\rvx_0 \equiv \rvx \sim p(\rvx)}$ into a Gaussian base distribution ${\rvx_{\mathcal{T}} \sim \mathcal{N}(\mathbf{0}, \rmI)}$. This continuous forward process ${\rvx_0\rightarrow\rvx_\tau \rightarrow\rvx_\mathcal{T}}$, with diffusion time $\tau\in\left[0, \mathcal{T}\right]$, can be trivially executed through adding Gaussian noise to initial data samples:
\begin{equation}
\label{eq:forward-diffusion}
    \rvx_\tau = \alpha_\tau \rvx_0 + \sigma_\tau \rvepsilon, \quad \rvepsilon \sim\mathcal{N}(\mathbf{0}, \rmI),
\end{equation}
where $\alpha_\tau$ and $\sigma_\tau$ are the signal and noise terms respectively, defined by some pre-defined noise schedule. Naturally, we are interested in the reversal of this corruption process, which is akin to sampling from the target distribution $p(\rvx)$. This can be interpreted as iteratively denoising the noisy estimate $\rvx_\tau$ ~\cite{milanfar2024denoising}. Tweedie's formula relates the \acrfull{mmse} to the score of the distribution:
\begin{equation}
\label{eq:tweedie}
    \rvx_{0\mid\tau} \defeq\, \mathbb{E}[\rvx_0\vert\rvx_\tau] = \frac{1}{\alpha_\tau}(\rvx_\tau + \sigma_\tau^2 \underbrace{\nabla_{\rvx_\tau} \log p(\rvx_\tau)}_{\text{score}}),
\end{equation}
where $\rvx_{0\mid\tau}$ represents the one-step denoised estimate from diffusion step $\tau$. The score function, defined as the gradient of the log-likelihood of $\rvx_\tau$, points towards the data distribution. Since Tweedie’s formula provides only a local (at time $\tau$) estimate, the final prediction of $\rvx_0$  is obtained iteratively. At each step, $\rvx_0$ is estimated using \eqref{eq:tweedie} and mapped back to $\rvx_{\tau-1}$ via forward diffusion in \eqref{eq:forward-diffusion}, ensuring a smooth sampling trajectory. The score function can be parameterized using a neural network, conditioned on the diffusion time step $\tau$. For practical reasons, the noise is predicted instead of the score, as it directly relates to the score via $\rvepsilon_\theta(\rvx_\tau, \tau)\approx -\sigma_\tau\nabla_{\rvx_\tau} \log p(\rvx_\tau)$ \cite{song2020denoising}. The network can be trained simply using the denoising score-matching objective as follows:
\begin{align}
    \mathcal{L}(\theta) = \E_{\rvx_0\sim p(\rvx_0), \rvepsilon, \tau} \left[ \left\| \rvepsilon_\theta(\rvx_\tau, \tau) - \rvepsilon \right\|^2 \right].
\label{eq:dsm}
\end{align}

\begin{figure*}
    \centering
    \includegraphics[width=0.8\linewidth]{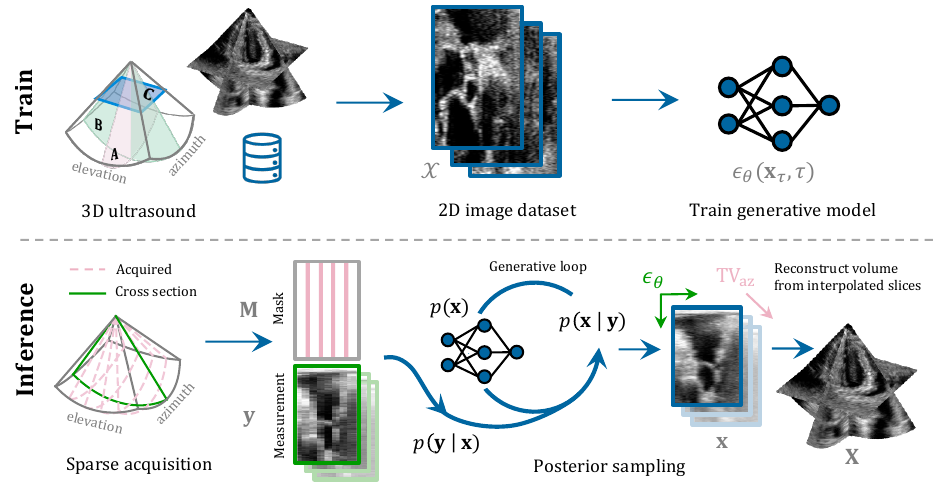}
    \caption{Overview of the training and inference pipeline for diffusion-based interpolation of 3D ultrasound volumes. A 2D score-based prior is learned from fully sampled elevation (B plane) slices extracted from the training dataset. At inference time, this generative prior guides the posterior sampling process to reconstruct from subsampled B planes. See Algorithm~\ref{algo:interpolation} for details of the proposed interpolation method using \acrshortpl{dm}.}
    \label{fig:overview}
    \vspace{-0.4cm}
\end{figure*}

\section{Methods}\label{sec:methods}
We formulate sparse 3D ultrasound reconstruction as an inverse problem, using \acrshortpl{dm} as image priors. This allows training a self-supervised model that learns the data distribution without paired examples. The learned prior is then applied during the inversion process, offering a task-agnostic framework that can be applied to various inverse problems without retraining. For the problem at hand, this flexibility allows for modifying the sampled scan-lines at test time, effectively changing the measurement matrix (i.e. sampling mask) adaptively. A conceptual overview of the proposed method is shown in Fig.~\ref{fig:overview}.

We begin by formally describing the inverse problem in \Secref{sec:3d-interpolation}, defining the interpolation task within a probabilistic framework and outlining how posterior sampling with diffusion models (see \Secref{sec:diffusion}) can be leveraged for reconstruction. Next, we discuss the choice of prior for 3D ultrasound data and detail the training process of the diffusion model in \Secref{sec:prior}. Finally, we explore strategies to incorporate temporal information during inference (\Secref{sec:temporal}) and assess the models' uncertainty in the interpolated results (\Secref{sec:uncertainty}).

\subsection{Interpolation of 3D ultrasound}\label{sec:3d-interpolation}
As discussed in \Secref{sec:3d-ultrasound}, the fully sampled 3D ultrasound volume is given by $\tX$, while the partially observed volume, following the Bayesian framework, is denoted by $\tY$. Using their vectorized forms, $\rvx$ and $\rvy$, respectively, the inverse problem of inpainting missing data can be formulated as:
\begin{equation}
    \rvy = \mA \rvx,
    \label{eq:inverse-problem}
\end{equation}
where the binary measurement matrix $\mA \in \{0,1\}^{(N_\text{el} / r) \times N_\text{el}}$ has an acceleration rate $r\geq 1$. The measurement matrix is related to the element-wise masking operator $\mM$ as follows:
\begin{equation}
    \mM = \operatorname{diag}(\mA^\top\mA)
\end{equation}
which allows us to express the backprojected (zero-filled) observations as:
\begin{equation}
    \rvy_{\text{zf}} =\mA^\top \rvy = \mM \odot \rvx,
    \label{eq:observation_zf}
\end{equation}
which we will use later primarily for the visualization of both measurements and the mask in the full image space.
Since \eqref{eq:inverse-problem} is underdetermined, inferring $\rvx$ from $\rvy$ is inherently challenging. There are many possible solutions that could explain the observed data. To address this, strong prior knowledge about the structure of $\rvx$ must be incorporated, guiding the reconstruction towards a plausible solution. This motivates the use of deep generative models, specifically \emph{posterior sampling} with diffusion models. Given the linear forward model in \eqref{eq:inverse-problem}, the posterior distribution $p(\rvx | \rvy)$ can be sampled through simple application of Bayes' rule for scores:
\begin{equation}
    \underbrace{\nabla_{\rvx_\tau} \log p(\rvx_\tau | \rvy)}_{\text{posterior}} = \underbrace{\nabla_{\rvx_\tau} \log p(\rvy | \rvx_\tau)}_{\text{likelihood}} + \underbrace{\nabla_{\rvx_\tau} \log p(\rvx_\tau)}_{\text{prior}}
    \label{eq:bayes-score}
\end{equation}
while the prior is modeled through the score network (see \Secref{sec:diffusion}), the likelihood term is closed-form. The likelihood term ensures that generated samples are consistent with the measurement and is also referred to as \emph{guidance} as it guides the diffusion process. This expression needs to be evaluated for all diffusion time steps $\tau$, which is typically intractable~\cite{daras2024survey}. There are several suitable posterior sampling methods for diffusion models, such as $\Pi$GDM\cite{song2023pseudoinverseguided}, RED-Diff~\cite{mardani2023variational}, which circumvent this issue. In this work, we will use the widely adopted \acrfull{dps} method~\cite{chung2022diffusion}, which offers a simple and effective gradient-based solution. \Acrshort{dps} interleaves prior updates (i.e. denoising) with guidance steps (gradient step towards the measurement). We now derive the corresponding expression for the time-dependent log-likelihood score used to guide the diffusion process via \acrshort{dps}, based on the formulation in \eqref{eq:bayes-score}. Starting in \eqref{eq:dps-linear-1}, we address the intractability of the noise-perturbed likelihood score $\nabla_{\rvx_\tau} \log p(\rvy | \rvx_\tau)$ by substituting $\rvx_\tau$ with Tweedie's estimate $\rvx_{0\mid\tau}$ from \eqref{eq:tweedie}, resulting in an analytically tractable formulation~\cite{chung2022diffusion, daras2024survey}:

\begin{align}
    \nabla_{\rvx_\tau} & \log p(\rvy | \rvx_\tau) \approx \nabla_{\rvx_\tau} \log p\left(\rvy \, | \, \rvx_{0\mid\tau} \right) \label{eq:dps-linear-1} \\
    &= -\frac{1}{2\sigma_{\rvn}^2} \nabla_{\rvx_\tau} \norm{\rvy - \mA \rvx_{0\mid\tau}}_2^2 \label{eq:dps-linear-2} \\
    &= \tsw{\frac{1}{\sigma_{\rvn}^2}} \nabla_{\rvx_\tau}^\top \rvx_{0\mid\tau} \mA^\top \left( \rvy - \mA \rvx_{0\mid\tau} \right)\label{eq:dps-linear-3}  \\
    &\approx - \underbrace{\gamma}_{\substack{\text{Guidance}\\\text{strength}}} \underbrace{\left(\mI - \sigma_\tau\nabla_{\rvx_\tau} \rvepsilon_\theta(\rvx_\tau, \tau) \right)^\top
        \mA^\top}_{\text{Projection}}
        \underbrace{\left( \rvy - \mA \rvx_{0\mid\tau} \right)}_{\substack{\text{Measurement}\\\text{error}}}. \label{eq:dps-linear-4}
\end{align}
In \eqref{eq:dps-linear-2}, we relax the delta likelihood, as follows from our noiseless forward model \eqref{eq:inverse-problem}, as a Gaussian $\mathcal{N}(\rvy; \mA \rvx, \sigma_{\rvn}^2 \mI)$. This leads to the intuitive $\ell_2$ norm between the observation and the denoised estimate $\rvx_{0\mid\tau}$ of the diffusion model.  Taking the gradient of the quadratic term yields \eqref{eq:dps-linear-3}, where the residual $(\rvy - \mA\rvx_{0\mid\tau})$ represents the measurement error, projected back into the signal space by $\mA^\top$. Finally, in \eqref{eq:dps-linear-4}, we use \eqref{eq:tweedie} again and subsequently substitute the actual score, which is used to compute $\rvx_{0\mid\tau}$, for the trained score model, introducing the final approximation. As is commonly done~\cite{daras2024survey}, we depart from the theoretical guidance strength by reweighting the likelihood with hyperparameter $\gamma$ in \eqref{eq:dps-linear-4}. An overview of the full algorithm for interpolating 3D ultrasound data, combining both diffusion (prior) and guidance (likelihood) steps, is shown in Algorithm~\ref{algo:interpolation}, also incorporating design choices detailed in the following two Sections~\ref{sec:prior} and~\ref{sec:temporal}.

\algnewcommand{\RequireOptional}{\item[\textbf{Optional:}]}
\algrenewcommand\algorithmiccomment[1]{\hfill\textcolor{gray!60}{$\triangleright$~#1}}
\newcommand{\CommentLine}[1]{%
  \Statex \hspace*{\ALG@thistlm} \textcolor{gray!60}{$\triangleright$~#1}%
}

\begin{algorithm}
\caption{3D ultrasound interpolation using \acrshortpl{dm}}
\begin{algorithmic}[1]
\Require partial volume $\tY \in \sR^{\tsw{N_\text{el}/r} \times N_\text{az} \times N_\text{ax}}$, subsampling rate \tsw{$r \geq 1$}, measurement matrix $\mA \in \{0,1\}^{\tsw{(N_\text{el} / r)} \times N_\text{el}}$, score model $\rvepsilon_\theta(\cdot)$, guidance strength $\gamma$, diffusion steps $\mathcal{T}$, noise schedule $\alpha_\tau, \sigma_\tau$ for $\tau \in [0, \mathcal{T}]$ (discretized), smoothness strength $\zeta$
\RequireOptional previous reconstruction $\tX^{\text{prev}}$, accelerated step $\tsw{\tau^\prime}$
\Ensure reconstructed volume $\tX_0 \in \sR^{N_\text{el} \times N_\text{az} \times N_\text{ax}}$

\CommentLine{Initialization}

  \If{$\tX^{\text{prev}}$ available}
    \State $\tX_0 \gets \tX^{\text{prev}}$ \Comment{SeqDiff start~\cite{stevens2024sequential}}
    \State $\tE \sim \mathcal{N}(\mathbf{0}, \tI)$
    \State $\tX_{\tau} \gets \alpha_{\tsw{\tau^\prime}} \tX_0 + \sigma_{\tsw{\tau^\prime}} \tE$ \Comment{Forward diffusion}
  \Else
    \State $\tX_\tau \gets \tX_{\mathcal{T}} \sim \mathcal{N}(\mathbf{0}, \sigma^2_{\mathcal{T}} \tI)$ \Comment{Cold start}
    \State $\tsw{\tau^\prime} \gets \mathcal{T}$
  \EndIf
  \CommentLine{Reverse (guided) diffusion loop}
  \For{$\tau = \tsw{\tau^\prime}$ to $0$}
    \For{all \textcolor{ForestGreen}{B planes} $\rvx_\tau, \rvy$ in $\tX_\tau, \tY$ (parallel)}
    \State $\rvepsilon \gets \rvepsilon_\theta(\rvx_\tau, \tau)$ \Comment{Predict noise}
    \State $\rvx_{0|\tau} \gets \frac{1}{\alpha_\tau} (\rvx_\tau - \sigma_\tau \rvepsilon)$ \Comment{Denoise (prior)}
    \State $\mathcal{M} \gets \rvy - \mA \rvx_{0|\tau}$ \Comment{Measurement error}
    \State $\mathcal{P} \gets \left( \rmI - \sigma_\tau \nabla_{\rvx_\tau} \rvepsilon_\theta(\rvx_\tau, \tau)\right)^T \mA^T$ \Comment{Projection}
    \State $\rvx_{0|\tau} \gets \rvx_{0|\tau} - \gamma \mathcal{P} \mathcal{M}$ \Comment{Guidance step}
    \State $\rvx_{\tau-1} \gets \alpha_{\tau-1} \rvx_{0|\tau} + \sigma_{\tau-1} \rvepsilon$ \Comment{Forward diffusion}
    \EndFor
    \State $\tX_{\tau - 1} \gets [ \rvx_{\tau - 1}^0, \ldots,  \rvx_{\tau - 1}^{N_{\text{el}}} ]$ \Comment{Stack planes}
    \State $\mathcal{V} \gets \nabla_{\tX_\tau}\text{TV}_{\text{\textcolor{Lavender}{az}}}(\tX_{\tau-1})$ \Comment{Total variation norm}
    \State $\tX_{\tau-1} \gets \tX_{\tau-1} - \alpha_{\tau-1} \zeta \mathcal{V}$ \Comment{Smoothness step}
    \EndFor
\Return $\tX_0$
\end{algorithmic}
\label{algo:interpolation}
\end{algorithm}

\subsection{Training and choice of prior}\label{sec:prior}
Selecting an appropriate representation for learning 3D ultrasound priors is nontrivial, particularly when considering all possible design options for how to present the data to the diffusion model. A seemingly straightforward approach is to employ a fully 3D generative model to learn the 3D prior; however, this introduces several challenges. First, as the dimensionality of both the data and model increases, the amount of training data required to capture the distribution grows exponentially~\cite{koppen2000curse}, making 3D generative models highly \emph{data hungry}. Second, 3D models impose significant computational demands, requiring substantial GPU memory and extended training times, which limits their feasibility in practical applications. At the same time, 3D ultrasound data exhibits strong spatial and temporal correlations, which can be leveraged to improve efficiency when transitioning from 2D to 3D representations. These considerations motivate our choice to adopt 2D models for learning a prior across the elevation dimension (B planes) of a 3D volume. To ensure consistent interpolation across the azimuth dimension (A planes), we employ an additional smoothness prior across this dimension using the total variation (TV) distance measure.
While prior work has demonstrated the success of 2D priors to high-dimensional data, including 3D MRI\cite{chung2023solving} and video data \cite{daras2024warped, stevens2024sequential}, they have not yet been explored for volumetric ultrasound. Additionally, 2D models and 2D ultrasound datasets are more readily available, making it easier to leverage pretrained models and integrate them into the proposed framework. Therefore, given the fully sampled 3D ultrasound volume $\tX$ as defined in \eqref{eq:ultrasound-volume}, we construct a dataset $\mathcal{X}~=~\left\{\rvx_{\text{el}}^{(1)}, \rvx_{\text{el}}^{(2)}, \ldots,  \rvx_\text{el}^{(|\mathcal{X}|)}\right\}~\sim~p(\rvx_\text{el})$, which is composed of elevation slices (\textcolor{ForestGreen}{B plane}). In relation to the full volume $\tX$, the slice is highlighted in \textcolor{ForestGreen}{green} as follows:
\begin{equation}
    \tX =
    \left[\begin{array}{@{\hspace{0.8em}}c@{\hspace{0.5em}}c@{\hspace{0.5em}}c@{\hspace{1em}}c@{\hspace{0.8em}}}
        \tikzmarkin[ver=style green]{firstcol2}\rvx_{1,1} & \rvx_{1,2} & \cdots & \rvx_{1,N_\text{az}} \\[0.4em]
        \rvx_{2,1} & \rvx_{2,2} & \cdots & \rvx_{2,N_\text{az}} \\
        \vdots & \vdots & \ddots & \vdots \\[0.4em]
        \rvx_{N_\text{el},1}\tikzmarkend{firstcol2} & \rvx_{N_\text{el},2} & \cdots & \rvx_{N_\text{el},N_\text{az}}
    \end{array}\right].
\end{equation}
\vspace{-1.5em}
\begin{equation*}
\underbrace{\hspace{2em}}_{\rvx_{\text{el}^{(1)}}}\hspace{7.2em}
\end{equation*}
Moreover, we use B-mode image data, as defined in \eqref{eq:ultrasound-volume}, in the native polar coordinate system with $N_\text{el}=48$, $N_\text{az}=64$, and $N_\text{ax}=400$. For visualization purposes, all images presented in this paper are shown after scan-conversion to a Cartesian grid, allowing the anatomy to be displayed in its true physical coordinate system.

Our dataset consists of 100 \emph{in vivo} volumetric cardiac cine-loops, recorded using an X5-1C matrix phased array transducer connected to a Philips EPIQ scanner. The volumes were collected from various sites and scanning sessions involving 16 patients and contain, on average, 40 frames spanning at least one full cardiac cycle. Approximately 10\% of the dataset is reserved for validation and testing, with one cine-loop designated for validation and hyperparameter tuning, and the remaining seven cine-loops from three patients used for testing.

The diffusion model is trained using the denoising score matching objective from \eqref{eq:dsm}, with a time-conditioned \mbox{U-Net} ($\approx$\SI{3.9}{\mega\nothing} parameters) and sinusoidal embeddings~\cite{pmlr-v70-gehring17a} as the backbone architecture. We optimize the model using AdamW with a learning rate of \num{1e-4} and weight decay of \num{1e-4}.] Finally, we turn on \acrfull{ema} of \num{0.999}, minimizing an MSE loss over approximately 25 epochs, using a batch size of 16. The data is clipped to a dynamic range of \SI{50}{\decibel} and normalized to the interval $[-1,\,1]$.

\subsection{Temporal consistency}\label{sec:temporal}
Ensuring temporal consistency across frames is crucial for high-quality 3D ultrasound reconstruction. By explicitly incorporating temporal information, we can produce more coherent and stable reconstructions over time, reducing flicker and improving clinical interpretability. To achieve this, we build upon \acrshort{seqdiff}~\cite{stevens2024sequential}, which exploits temporal continuity by initializing the current frame’s diffusion process from the previous frame’s reconstruction. Rather than starting the generative process from pure noise at the beginning of each frame, \acrshort{seqdiff} warm-starts the diffusion at an intermediate step along the reverse trajectory, effectively integrating temporal consistency into the sampling process. Formally, \Acrshort{seqdiff} initializes the diffusion process at some intermediate step $\tau^\prime$ along the reverse diffusion trajectory, where $0 < \tau^\prime \ll \mathcal{T}$, effectively reducing the number of iterations. Rather than beginning each generation from scratch at $\tau = \mathcal{T}$ using a pure Gaussian sample $\rvx_\mathcal{T} \sim \mathcal{N}(\mathbf{0}, \sigma^2_\mathcal{T} \mI)$, \acrshort{seqdiff} initializes the process using a previous solution $\rvx_0^{t-1}$, forward-diffused to the current $\tau^\prime$.

Specifically, we reconstruct from a sequence of time-dependent measurements $\rvy^t = \mA^t \rvx^t$, where $\mA$ varies according to the sparse interlocking acquisition scheme as described in Fig.~\ref{fig:acquisition-scheme} and $\rvx^t$ and $\rvx^{t-1}$ are assumed to exhibit strong structural similarity. By propagating prior reconstructions forward to an intermediate noise level and using them as initialization, we significantly reduce computation and improve temporal consistency in the reconstructions. This approach is well-suited for cardiac ultrasound, where, at high frame rates, the anatomy changes gradually over time. Thereby, the previous frame’s reconstruction serves as an effective prior, allowing each newly acquired set of elevation planes to benefit from earlier information. Thus, $\tau^\prime$ is chosen as low as possible to preserve temporal context without hindering reconstruction of current measurements. Following \cite{stevens2024sequential} we set it to $\approx 20\%$ of the total diffusion steps.

\begin{figure*}
    \centering
    \begin{tikzpicture}
        \node[anchor=south west, inner sep=0] (image) at (0,0)
            {\includegraphics[width=0.95\linewidth, trim=0 0 0 0, clip]{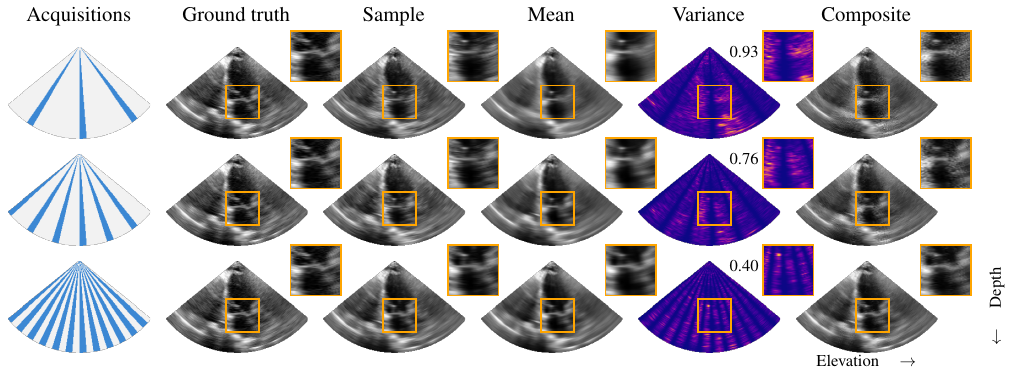}};
        \begin{scope}[x={(image.south east)}, y={(image.north west)}]
            \node[rotate=90, text={rgb,255:red,111;green,106;blue,106}] at (-0.01, 0.19) {$2\times$};
            \node[rotate=90, text={rgb,255:red,111;green,106;blue,106}] at (-0.01, 0.47) {$4\times$};
            \node[rotate=90, text={rgb,255:red,111;green,106;blue,106}] at (-0.01, 0.78) {$8\times$};
        \end{scope}
    \end{tikzpicture}
    \caption{A single posterior sample, the posterior mean and variance over multiple samples, and the composite images for reconstructions with fractions of the elevation planes acquired \mbox{$r = (8\times, 4\times, 2\times)$}. It is clear that the posterior samples vary more in the unmeasured regions of tissue, and that acquiring more elevation planes reduces the overall uncertainty and boosts the reconstruction quality. This is also reflected in the entropy measure $H(\rvx)$ displayed in the corner of each variance image.}
    \label{fig:uncertainty}
\end{figure*}

\subsection{Uncertainty quantification}\label{sec:uncertainty}
Many interpolation methods provide only a single \emph{point-estimate} reconstruction, failing to represent the uncertainty inherent to the inverse problem. Given the probabilistic nature of \acrshort{dps}, however, our method can quantify the uncertainty present in its reconstructions. Quantifying the uncertainty can help mitigate \textit{hallucinations}, in which a model generates one of many plausible anatomies that may be convincing but not truthful. Or, more critically, it overlooks an anatomical feature that is, in fact, present. In what follows, we describe two methods to quantify and communicate this uncertainty. We build on the well-established idea of combining samples from the posterior distribution~\cite{ekmekci2023quantifying, adler2024deep}. Our contribution lies in adapting this principle to 3D ultrasound reconstruction and proposing visualization strategies that convey the uncertainty in an interpretable manner to the operator.

As discussed in \Secref{sec:3d-interpolation}, the posterior sampling approach allows us to draw $N$ independent samples $\{\rvx^{(i)}\}_{i=0}^N~\sim~p(\rvx|\rvy)$ from the posterior, where each $\rvx^{(i)}$ is obtained after a completed reverse diffusion process. Computing statistics across these samples provides uncertainty estimates that reflect variability in the final reconstructions. Luckily, drawing multiple posterior samples can be executed in parallel on the GPU, minimizing the additional computational expense. We propose two methods for visualizing the model's uncertainty, along with an outlook for future work. An example of both methods for various acceleration rates is shown in Fig.~\ref{fig:uncertainty}.

\subsubsection{Variance heat map} By analyzing the pixel-wise variance across posterior samples, we can assess which regions exhibit higher or lower certainty, where $\bar{\rvx}$ is the sample mean:
\begin{equation}
\sigma_\rvx^2 = \mathbb{E}\left[ \left( \rvx - \mathbb{E}[\rvx] \right)^2 \right] \approx \frac{1}{N}\sum_i\left[ (\rvx^{(i)} - \bar{\rvx})^2\right].
\end{equation}
For instance, directly acquired scan lines are well-constrained by the measurement model and therefore exhibit minimal variance. In contrast, non-acquired scan lines rely on the generative model for reconstruction, where higher uncertainty manifests as greater variance in the posterior samples. To obtain a scalar measure of uncertainty for each reconstructed volume, we compute the entropy under a multi-variate Gaussian model with diagonal covariance of the posterior as follows:
\begin{equation}
    H(\rvx) = \frac{1}{2N} \sum_j \ln(2\pi e \sigma_{\rvx, j}^2),
\end{equation}
where $\sigma_{\rvx, j}^2$ denotes the variance of the $j$'th pixel in $\rvx$. As shown in Fig.~\ref{fig:uncertainty}, the entropy increases consistently with higher acceleration rates, reflecting the expected growth in uncertainty.

\subsubsection{Composite image} An alternative approach to uncertainty visualization involves creating a composite image by sampling pixels randomly from each posterior sample.
\begin{equation}
    \forall j\in\big[0, \text{len}(\rvx)\big): \tilde{\rvx}_j = \rvx^{(i)}_j, \quad \text{with } i \sim \text{Uniform}(1, N)
\end{equation}
This introduces perceivable noise in uncertain regions while maintaining a consistent appearance in regions where the samples agree. The rationale for this presentation is that, intuitively, the operator will naturally place less confidence in the regions that appear noisy, as noisier areas indicate greater model uncertainty.

Finally, beyond visualization techniques, it is crucial to recognize the inherent limitations of the model. In particular, it is important to consider that, in unobserved regions, the posterior samples will only vary according to the distribution that has been learned by the model, and potentially fail to include valid reconstructions that fall outside the training distribution. It is therefore  essential to understand how much subsampling remains acceptable for a given application. In the following experiments section, we analyze the model’s performance across various tasks and acquisition rates to determine the trade-offs between image quality and subsampling levels.

\section{Experiments}
We perform both qualitative and quantitative analyses to assess the benefits of advanced interpolation methods for 3D ultrasound reconstruction. First, we investigate image quality directly through the use of common image quality metrics. We leverage both (pixel-wise) distortion and (global) perceptual metrics. Finally, we perform an experiment on speckle tracking, a common clinical downstream task in cardiac imaging, to further validate our approach. All results are generated with DDIM sampling strategy~\cite{song2020denoising}, cosine noise schedule~\cite{nichol2021improved}, and parameters $\mathcal{T}=200, \tau^\prime=50, \gamma=35, \zeta=0.001$, which were optimized on the validation split and evaluated on the test split as outlined in \Secref{sec:prior}. We implemented the proposed method using the zea ultrasound toolbox \cite{zea2025}, with JAX as the compute backend.

\subsection{Baselines}
In every experiment, we consider the fully sampled volume as the ground truth. This allows for both quantitative evaluation of reconstruction quality and comparison of downstream task performance between interpolated and fully acquired volumes. Image quality metrics are computed in the original polar domain, where interpolation is performed to avoid confounding effects from scan conversion, which itself applies interpolation.

\subsubsection{Linear and nearest-neighbor interpolation}
The most straightforward methods for volumetric interpolation in 3D ultrasound are nearest-neighbor and linear interpolation along the elevation axis. We include these traditional techniques as simple baselines for reference. Nearest-neighbor fills missing slices by duplicating the nearest acquired slice, effectively visualizing the sparse measurements in the full volume. This approach introduces blocky artifacts and discontinuities. Linear interpolation averages neighboring slices to produce smoother transitions, but it often results in blurring and loss of detail. Both methods are low in complexity and can be easily implemented in clinical systems. However, they lack anatomical awareness, which limits their effectiveness at higher acceleration rates. While they may suffice at lower rates, doing so constrains the potential temporal resolution gains targeted in this work.

\subsubsection{End-to-end supervised}
Additionally, we include a baseline that trains a deep learning network in a fully supervised, end-to-end manner for inpainting 3D ultrasound. Specifically, we employ the same \mbox{U-Net} architecture as used by the diffusion model of the proposed approach (see \Secref{sec:prior}), and focus on the inpainting of individual elevation slices, maintaining the same problem scope. Unlike the diffusion model, which leverages an explicit forward model during sampling, this baseline does not explicitly incorporate forward model knowledge during inference. Instead, it learns a data-driven mapping $\rvx = f_\theta(\rvy_{\text{zf}})$ from zero-filled input measurements $\rvy_{\text{zf}}$, as seen in \eqref{eq:observation_zf}, to complete reconstructions. Consequently, the network’s performance may degrade on unseen mask patterns, in contrast to the diffusion model’s task-agnostic approach.

\subsubsection{Neural field}
Finally, we include a neural field, also known as an \acrfull{inr}, which parameterizes a continuous volumetric field with a neural network. A six-layer fully connected (dense) network $f_\theta : \R^4 \rightarrow \R$, with similar parameter count as the other baselines, is fitted per volume to map continuous 4D coordinates $(t, i, j, k)$ to observed grayscale B-mode values. To improve representation capacity, sinusoidal positional embeddings~\cite{pmlr-v70-gehring17a} are applied to the input coordinates as commonly done~\cite{mildenhall2021nerf}. The network is trained on the entire 4D tensor to exploit all available spatiotemporal information. Once trained, the \acrshort{inr} can be evaluated at any spatial location, enabling interpolation of the data at arbitrary resolutions. \Acrshortpl{inr} have previously been applied to ultrasound for efficient storage of 3D volumes~\cite{gu2022representing} and volumetric reconstruction of freehand acquisitions~\cite{yeung2024sensorless}. However, reconstruction quality is limited by the network’s smooth architectural prior and its lack of awareness beyond the observed slices. In contrast, our generative prior is trained on fully acquired volumes, capturing realistic anatomical and speckle statistics. Furthermore, inference of \acrshortpl{inr} is slow due to the need for per-volume fitting prior to the interpolation.

\begin{figure*}
    \centering
    \includegraphics[width=1.0\linewidth]{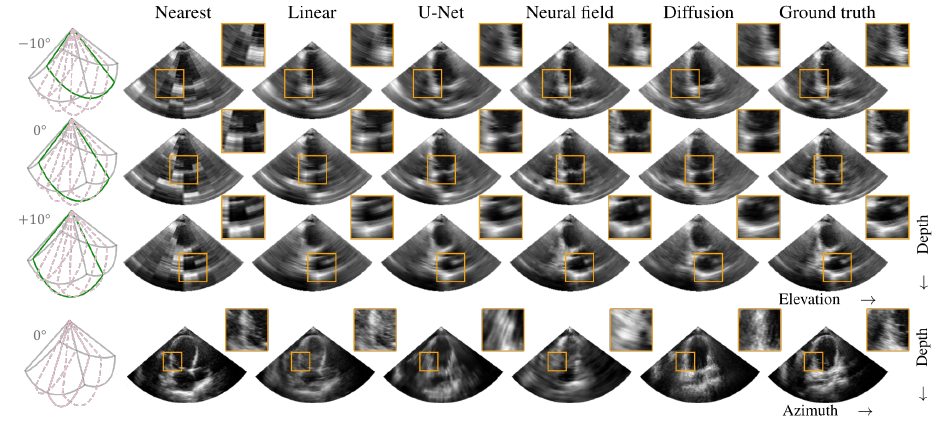}
    \caption{Qualitative comparison of the proposed diffusion-based interpolation method with baseline approaches (nearest, linear, U-Net, and neural field). The figure displays three \textcolor{ForestGreen}{B planes} at various azimuth angles and an \textcolor{Lavender}{A plane}, from a single 3D volume, with their positions within the frustum geometry illustrated on the left. While dashed slices represent the acquired planes, the solid lines denote the interpolated data. Acceleration rate is set to $r=3$. Additionally, zoomed insets are shown in the top right corner of each subplot.}
    \label{fig:qualitative}
\end{figure*}

\subsection{Image quality}
\begin{figure}
    \centering
    \includegraphics[width=0.9\linewidth]{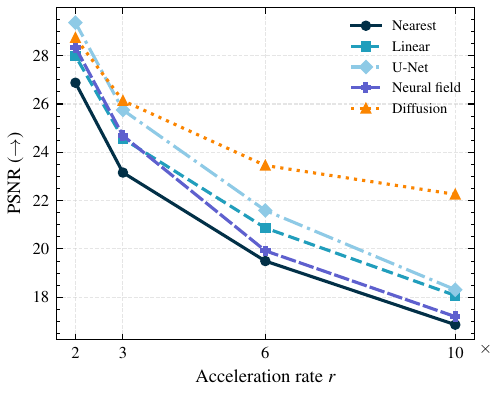}
    \caption{Reconstruction performance evaluated using a distortion metric (PSNR) across various interpolation methods at increasing acceleration rates \mbox{$r \in \{2, 3, 6, 10\}$}.}
    \label{fig:psnr-subsampling}
\end{figure}

\begin{figure}
    \centering
    \includegraphics[width=0.9\linewidth]{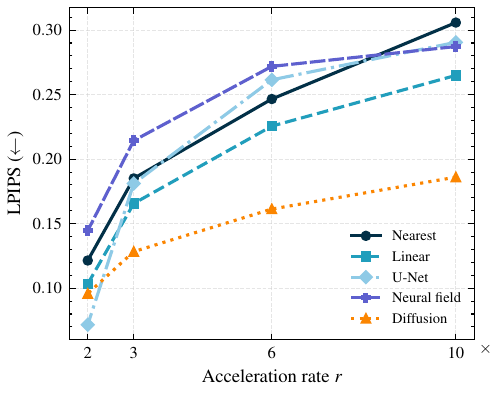}
    \caption{Reconstruction performance evaluated using a perceptual metric (LPIPS) across \textcolor{ForestGreen}{B planes} for various interpolation methods at increasing acceleration rates \mbox{$r \in \{2, 3, 6, 10\}$}.}
    \label{fig:lpips-subsampling}
\end{figure}

\begin{figure}
    \centering
    \includegraphics[width=0.9\linewidth]{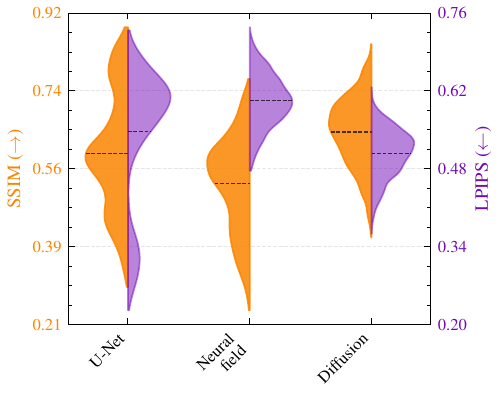}
    \caption{Image quality for the data-driven 3D interpolation methods. Testing interpolation performance across \textcolor{Lavender}{A planes} using (\acrshort{ssim}) (\textuparrow{}) (orange) and perceptual (\acrshort{lpips}) (\textdownarrow{}) (purple) metrics. Acceleration rate is set to $r=6$.}
    \label{fig:A-plane-image-quality}
\end{figure}

Fig.~\ref{fig:qualitative} shows a visual comparison of the proposed interpolation method with respect to the baselines. Additionally, we evaluate the interpolated volumes using \acrfull{psnr} as a pixel-wise distortion metric, as well as \acrfull{ssim} and \acrfull{lpips}~\cite{zhang2018perceptual} to measure perceptual quality, assessing both aspects of the perception--distortion trade-off \cite{blau2018perception}. While the distortion metric is calculated on the volume as a whole, the perceptual metrics are calculated on either A or B planes separately as \acrshort{lpips} and \acrshort{ssim} are implemented as a 2D image metrics.
Fig.~\ref{fig:psnr-subsampling} and Fig.~\ref{fig:lpips-subsampling} show the distortion and perceptual image quality performance against various acceleration factors $r$. Fig.~\ref{fig:A-plane-image-quality} shows the perceptual performance across A planes.
While the methods perform comparably well at $r=2$, with very low \acrshort{lpips} values indicating minimal perceptual differences, clear distinctions emerge at higher acceleration rates. The diffusion-based approach maintains strong perceptual and distortion metrics at higher acceleration rates, whereas baseline methods exhibit increasingly visible interpolation artifacts, as reflected by the metrics.

\subsection{Speckle tracking}
Speckle tracking is used to assess tissue motion and deformation by tracking speckle patterns across frames \cite{garcia2018introduction}, e.g., enabling strain measurements for evaluating myocardial function. While 2D speckle tracking is commonly used, it is limited by out-of-plane motion and foreshortening artifacts. Volumetric ultrasound imaging mitigates these limitations by providing full 3D motion tracking, thereby improving robustness and accuracy.

\begin{figure}
    \centering
    \includegraphics[width=0.9\linewidth]{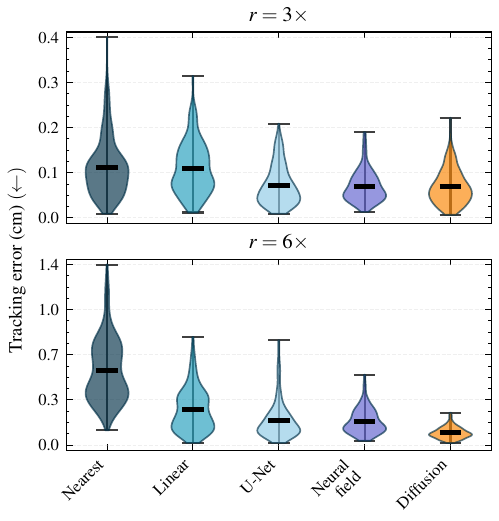}
    \caption{Speckle tracking performance comparison across various interpolation methods at acceleration rates of $r = 3, 6$. The proposed method shows significant improvement over baseline approaches at higher acceleration rates.}
    \label{fig:speckle-tracking-results}
\end{figure}

To assess the impact of different reconstruction methods on motion estimation, we evaluate 3D speckle tracking performance on both ground truth and interpolated volumes, comparing the resulting tracks to those obtained from the ground-truth data to quantify tracking error. Tracks are initialized automatically using a segmentation model that outlines the left ventricle~\cite{ouyang2020video}. Local tissue displacements between consecutive 3D frames are then estimated using the Lucas–Kanade optical flow method~\cite{leung2011left}, which determines voxel-wise motion by minimizing intensity differences within local neighborhoods. We perform tracking in the Cartesian domain with a window size of 32 voxels and three pyramid levels. We use the zea implementation for segmentation and tracking methods~\cite{zea2025}.
As shown in Fig.~\ref{fig:speckle-tracking-results}, the proposed method demonstrates a noticeable reduction in tracking error compared to the other approaches. This can be attributed to the integrated temporal consistency (\acrshort{seqdiff}) of the proposed method, as well as improved fidelity of the reconstructions.

\begin{figure*}
    \centering
    \subfloat{%
        \begin{tikzpicture}
            \node[anchor=south west, inner sep=0] (imgA) at (0,0) {\includegraphics[height=0.28\textwidth]{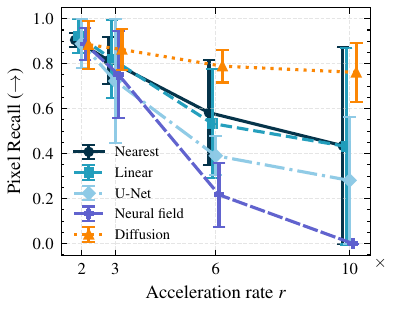}};
            \node[anchor=south west, font=\bfseries] at (imgA.south west) {(a)};
        \end{tikzpicture}
    }
    \hspace{0.0\textwidth}
    \subfloat{%
        \begin{tikzpicture}
            \node[anchor=south west, inner sep=0] (imgB) at (0,0) {\includegraphics[height=0.28\textwidth]{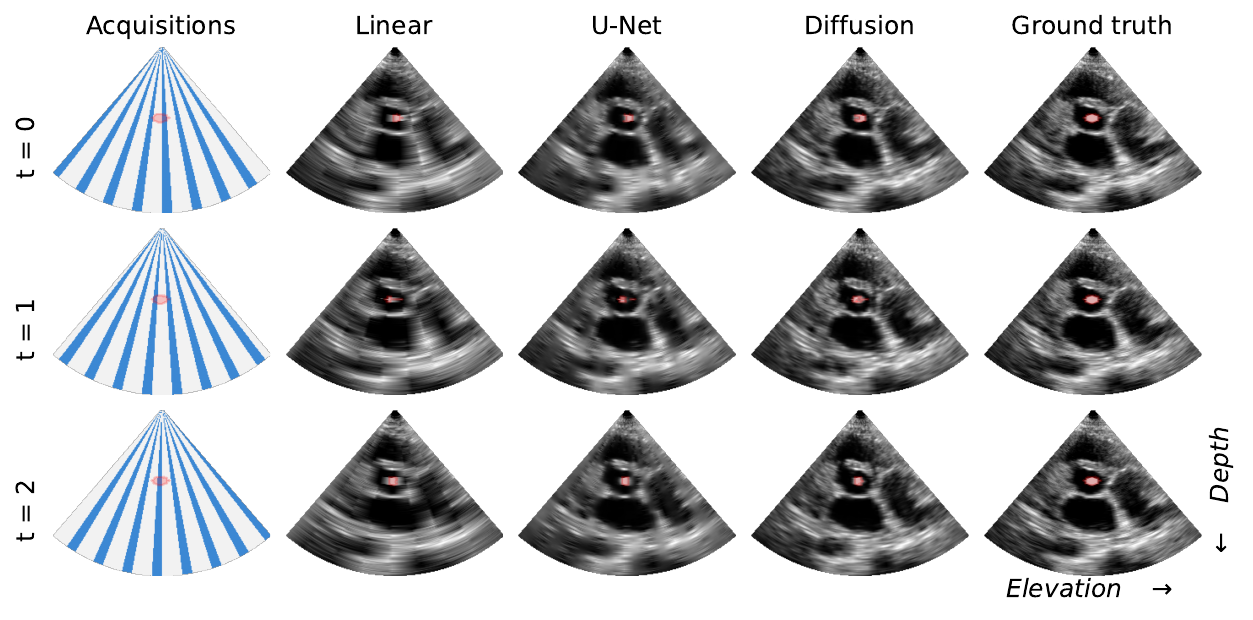}};
            \node[anchor=south west, font=\bfseries] at (imgB.south west) {(b)};
        \end{tikzpicture}
    }
    \vspace{-0.4\baselineskip}
    \caption{
        \textbf{(a)} Pixel recall for synthetic out-of-distribution inclusions at varying acceleration rates \mbox{$r \in \{2, 3, 6, 10\}$}, averaged over 10 randomly inserted inclusions with 95\% confidence intervals.
        \textbf{(b)} Example of three consecutive frames ($r=3$), with random inclusions highlighted in red \protect\redcircle{}. When the inclusion is not fully covered by an acquisition, baseline methods recover less of the circle, whereas the proposed method achieves better reconstruction due to built-in temporal consistency.
    }
    \label{fig:inclusion}
\end{figure*}

\subsection{Out-of-distribution data}
\label{sec:ood}
Our final experiment evaluates the robustness of the proposed method to \emph{out-of-distribution} (OoD) data. As discussed in \Secref{sec:uncertainty}, a concern with generative models is their tendency to \emph{hallucinate} structures or overlook important anatomical features when faced with ambiguous or unseen data. This experiment specifically investigates such failure modes by introducing synthetic anomalies into the test set.

We augment the test volumes with bright masses, which are simulated as circular inclusions of fixed diameter $5 \mathrm{px}$ and uniform brightness (70\%), randomly positioned within the volume. These inclusions are clearly \acrshort{ood}, as perfectly circular, high-contrast structures are not present in the training data. Importantly, this setup provides access to ground truth inclusion locations, allowing for quantitative evaluation. After subsampling and reconstruction, we compute the pixel-level recall of the inserted inclusions. A pixel is considered correctly recalled if its intensity lies within 10\% of the true inclusion value. The results are averaged over 10 randomly placed inclusions per configuration. See Fig.~\ref{fig:inclusion} for the recall performance across acceleration rates, as well as a visual example.

Interestingly, our method demonstrates consistently higher recall than baseline approaches, especially under strong subsampling. We attribute this to two key factors. First, the explicit conditioning on measured data during inference ensures consistency with the acquired measurements, improving reliability under distribution shifts. This contrasts with the supervised baselines trained end-to-end without an explicit measurement model. Second, by initialization of each reconstruction with the previous frame via \acrshort{seqdiff}, we do not only accelerate inference but also enhance robustness to anomalies by leveraging temporal continuity in the data.

\subsection{Acquisition and reconstruction rates}
\label{sec:speed}
When discussing the maximum achievable volume rate, it is important to distinguish between two separate factors: the \emph{acquisition rate} and the \emph{reconstruction rate}. The \emph{acquisition rate} is determined solely by the total round-trip travel time of all transmit events, independent of the reconstruction method. If reliable reconstruction can be achieved using fewer focused transmits across elevation, the acquisition rate can be effectively increased. The \emph{reconstruction rate}, in contrast, is directly governed by how fast the acquired data can be interpolated to reconstruct a full 3D volume, which is highly dependent on the specific algorithm used.

\textbf{Acquisition:}
For an imaging depth of \SI{20}{\centi\metre} and a typical speed of sound in tissue of \SI{1540}{\metre/\second}, the time per transmit is approximately \SI{260}{\micro\second}. Using 400 diverging waves to acquire a full 3D volume yields an acquisition rate of roughly \SI{10}{volumes/\second}. An acceleration rate of $r=3$, would reduce the required transmits and thus increase the acquisition rate by $3\times$, resulting in \SI{30}{volumes/\second}.

\textbf{Reconstruction:}
For reference, reconstruction rates for the baselines are \SI{2}{\second} (nearest), \SI{3}{\second} (linear), \SI{0.4}{\second} (supervised), \SI{13}{\second} (neural field) per volume, respectively.
The current implementation of the proposed algorithm takes about \SI{2}{\second/volume} at 50 diffusion steps on an NVIDIA L40s (48GB of VRAM). Notably, the reconstruction of a single B plane takes \mbox{$\sim\,$\SI{80}{\milli\second}}. The algorithm is designed to be fully parallelizable within each volume (across slices), suggesting that future optimization of the implementation and improvements in hardware could substantially improve this reconstruction rate, opening the door to real-time inference. In its current form, the approach is well-suited for offline 3D volume reconstruction, as operators can navigate using the standard 2D view, based on the fully acquired A plane, during scanning~\cite{huang2017review}, and generate fully interpolated volumes post-acquisition for detailed clinical review.

\section{Discussion}
This work demonstrates the effectiveness of \acrshort{dm}-based interpolation for reconstructing volumetric ultrasound from sparsely acquired elevation planes. This section outlines key implications of our results, including benefits of the probabilistic formulation, temporal consistency, and potential for real-time deployment. We also discuss limitations and future directions for integration into clinical workflows and future adaptive imaging systems.

A key strength of our method lies in its probabilistic formulation. The posterior sampling framework not only enhances robustness to \acrshort{ood} data but also yields uncertainty estimates that can inform downstream tasks or guide acquisition. This is especially relevant for integration with active imaging systems~\cite{federici2024active, van2024active, nolan2024active}, where uncertainty can drive intelligent transmit selection to minimize ambiguity in real time. In this work, we focus on reconstruction from a fixed acquisition pattern, but our approach lays important groundwork for closed-loop, adaptive imaging pipelines.

Temporal consistency is another benefit, achieved by incorporating \acrshort{seqdiff} into the sampling process. The integration of prior information stabilizes the generative process over time, enabling more coherent predictions and reducing redundant computations. In practice, we observe it improves both perceptual quality and robustness under aggressive subsampling.

A critical aspect concerns the feasibility of real-time deployment. From a practical standpoint, the approach achieves increased acquisition rates and, owing to its inherent parallelizability, shows strong potential for high-throughput reconstruction. Further optimization, through improved implementation, future hardware acceleration, or algorithmic refinements such as model distillation~\cite{meng2023distillation}, latent diffusion models~\cite{rombach2022high}, or advanced sampling strategies~\cite{ma2024deepcache} is required to substantially reduce inference time for real-time deployment.

In the absence of real-time operation, the method is well suited for offline clinical workflows. In many cases, only a 2D view is presented on the display, even with 3D imaging turned on~\cite{huang2017review}. Similarly, applications like patch ultrasound~\cite{hu2023wearable} do not require any real-time imaging. In such cases, low-resolution previews can guide acquisition, while high-quality reconstructions are generated afterward for diagnostic interpretation. This deferred processing model aligns with existing practices in the clinical workflow, where clinicians often analyze stored image sequences after a scanning session.

Beyond computational aspects, generalization of generative models remains an important direction. The model was trained and evaluated on a single \emph{in vivo} dataset acquired with one imaging system, which naturally limits our ability to assess performance across different anatomies, transducers, or institutions. The out-of-distribution experiment provides a controlled evaluation under domain shift by introducing synthetic inclusions, yet future studies on anatomically diverse datasets will be essential to fully characterize robustness and facilitate clinical translation.

\section{Conclusion}
3D ultrasound holds significant promise for the future of ultrasound imaging, particularly in cardiac applications. However, its clinical adoption remains limited due to constraints in temporal and spatial resolution, often leading clinicians to fall back to 2D imaging. To push this trade-off, one can employ spatial subsampling to reduce the number of transmit events, thereby enabling higher frame rates or broader coverage. Yet, conventional interpolation methods, as well as supervised learning-based approaches, struggle to fully reconstruct undersampled volumes, especially at high acceleration rates. In this work, we reframe 3D ultrasound reconstruction as a Bayesian inverse problem, leveraging deep generative priors to substantially improve reconstruction quality. Our approach demonstrates clear benefits not only in terms of visual fidelity but also in clinically relevant downstream tasks such as speckle tracking. Moreover, by extending our diffusion-based interpolation framework to exploit the sequential nature of ultrasound data, we achieve both improved temporal consistency and significant acceleration, bringing the method closer to real-time applicability. Finally, the probabilistic nature of our model enables estimation of reconstruction uncertainty, offering a valuable tool for clinical decision-making and a foundation for future work on uncertainty-aware imaging and adaptive acquisition strategies~\cite{van2024active}.

\bibliography{references}

\end{document}